\documentstyle[12pt]{article}
\begin{document}

\begin{center}

{\bf \large{ SPATIOTEMPORAL CHAOS OF SOLITON IN A GENERALIZED SKYRME MODEL WITH THE MODIFIED  SYMMETRY - BREAKING TERM . }}

\bigskip

{\bf Nguyen Vien Tho }\footnote{E-mail: ngvtho@dng.vnn.vn}

{\it Hue  University, Hue, Vietnam}

{\bf Phu Chi Hoa}\footnote{E-mail: pchihoa@hcm.vnn.vn}

{\it Dalat University, Dalat, Vietnam}

\end{center}

\vspace{0.5cm}

\begin{abstract}

 The chiral symmetry-breaking term of the Skyrme model with massive pion is modified to obtain the hedgehog profile function which is in best coincidence with the kink-like profile function. For the modified Lagrangian, the minimum of the energy of the B=2 twisty skyrmion configuration is lower than the values for both the cases of the Skyrme Lagrangian with and without the non-modified symmetry-breaking term. The equations  of motion for the time-dependent hedgehog of  this model and for a generalizated Skyrme model including sixth-order stabilizing term  are  derived and integrated nummerically. The time evolution of soliton is obtained. We have observed the seft-exitation of soliton because of the fast developement of fluctuation.

\end{abstract}

\bigskip

\noindent
{{\bf I.  INTRODUCTION}}

\vspace{0.2cm}

       It  is   known  that   one
can       give    a
description of low  energy   hadron
physics        on   the base of
a semiclassical  quatization of the soliton solution of
the  Skyrme model     
(skyrmion) $\left[ 1-3\right] .\,\,$The skyrmion configuration,  denoted
by $U (\overrightarrow{x})$, is a map:
$R^3\rightarrow SU(2)$, with the
condition $U\rightarrow 1$ as $\left| \overrightarrow{x}\right| \rightarrow
\infty$,
required        for  
finite     energy .   
One        could    
analyze   in details the case     of 
the  
spherically      symmetry 
configuration,  when   the matrix    
field  $ U (\overrightarrow{x})$ has 
the
form (the hedgehog):

$$
U(\overrightarrow{x})=\exp [\,\,i\,\,F(r)\overrightarrow{\,x}.\vec \tau ]
\eqno(1.1)
$$
where   $ r = \left| \overrightarrow{x}\right| $ and $\vec \tau$ is the Pauli    
matrices.    Here   $ F(r)$ is a profile
function
with the    boundary conditions  $F(0) = n\pi \,$ (n is an integer) 
and   F(r) $\rightarrow 0\,$as r$\rightarrow \infty $. F(r) obeys a
nonlinear differential      equation 
which    could   
be  solve     numerically.  

  An   analytic form of the
skyrmion profile function 
seems to be very      useful for many purposes.
Attemps    were   made to
obtain a such analytic approximation: 
(i) by
computing   the  holonomy   of  $SU(2)$
Yang- Mills    instantons in $R^4$ along lines
paralell to the time -axis $\left[4\right] $;
(ii) by identifying the    skyrmion profile
function
with     the sin-Gordon kink field 
(if we replace $x$ by $r$) $\left[ 5\right] $.
The approach (ii) is very attractive because of its simplicity.
Moreover,
unlike    the   instanton  
approach,    the   sin-Gordon
kink   has  a fixed   
scale,    so
there    are  no  abitrary 
scale  parameters      which 
have    to    be 
fixed    by     
hand in  order
to minimize the energy.
With   the explicit
expression of the kink-like profile function $F(r)$, in Sec.2,  the symmetry-breaking term of the Lagrangian is modified to obtain the hedgehog profile function for B=1 skyrmion which is in best coincidence with the kink-like profile function.  Then  we have calculated the energy of four different B=2 skyrmion configurations  and lead to the result: the lowest value of the energy is obtained in the case of twisty configuration for the modified Lagrangian [6].  

 The results on numerical integration of the equation of motion for the time-dependent hedgehog indicate a dynamical chaostic character of fluctuations around the static soliton solution [7]. Such a study is interesting from the viewpoint of considering the Skyrme model as a nonlinear dynamical system.  In Sec.3, we find the equation of motion for the time-dependent hedgehog of the considered model. The results of numerical integration of the obtained equation are presented. The fluctuation of the profile function $\delta F(x,t)$  as well as the time dependence of the amplitudes of different modes of the fluctuations are plotted. 

The Skyrme's  Lagrangian [1] is of fourth order in field derivatives. Various alternative models have been proposed which preserved the form of the original Lagrangian while extending it to higher orders [8-12]. The incorporation of the higher order terms, on the one hand, improves the fit of observables, and, on the other hand, gives a reasonable physical interpretation for stabilizing terms in Lagrngian.  For example, one could introduce a sixth-order term [8-9]

$$
{\cal L}^{\left( 6\right) }=-\frac{\varepsilon _6^2}2B^\mu B_\mu, 
\eqno(1.2)
$$
where  $ B^\mu$ is the baryon current

$$
B^\mu =\frac{\varepsilon ^{\mu \nu \alpha \beta }}{24\pi ^2}Tr[(U^{+}\partial _\nu U)(U^{+}\partial _\alpha U)(U^{+}\partial _\beta U)].
\eqno(1.3)
$$
This term may be understood an representing effects of  
$\varpi $ mesons, while the Skyrme's fourth-order term should be viewed as representing effects of  $\rho $ mesons. In Sec. 4, we consider the dynamical behavior  of the  model which the Skyrme Lagrangian added by the sixth-order term (1.2 )  and the  modified symmetry-breaking term. 
 The discussion of the results is given in Conclusion.

\bigskip

\noindent
{\bf II.  THE SIN-GORDON KINK FIELD AND THE CHOICE OF THE CHIRAL SYMMETRY BREAKING TERM}


\vspace{0.2cm}

	The Skyrme's Lagrangian density takes the form [2,3]
$$
{\cal L}=\frac{F_\pi ^2}{16}Tr\left( \partial _\mu U\partial ^\mu U^{+}\right) +\frac 1{32e^2}Tr\left[ (\partial _\mu U)U^{+},(\partial _\nu U)U^{+}\right] ^2.
\eqno(2.1)
$$

One considers also the Skyrme's Lagrangian with a pion mass term [3]
$$
{\cal L}^{\prime }={\cal L}+{\cal L}_{SB},
\eqno(2.2)
$$
$$
{\cal L}_{SB}= \frac 18m_\pi ^2F_\pi ^2Tr\left( U+U^{+}-2\right). 
\eqno(2.3)
$$

From (2.1), based on equations of motion,  one found the nonlinear differential equation for the B=1 hedgehog profile function
$$
\frac{xF^{\prime }}2+\left[ \frac{x^2}4+2Sin^2F\right] F^{^{\prime \prime }}+Sin2F\left( F^{^{\prime }}\right) ^2-\frac{Sin\left( 2F\right) }4-\frac{Sin^2FSin\left( 2F\right) }{x^2}=0,
\eqno(2.4)
$$
where $x=eF_\pi r$ is the dimensionless radial distance. 

 The kink-like function [5] has the form
$$
F\left( x \right) =4arctan \left( e^{-x}\right),
\eqno(2.5)
$$
which satisfies the same boundary conditions. The kink-like profile function (2.5) has an exponential decay for large x [5]. The same of an asymptotic behaviour is also obtained when the Skyrme Lagrangian (2.1) is added by a chiral symmetry-breaking term as (2.3) [13]. Accordingly, the right-hand side of (2.4) is not zero, but equals to 
$$
\frac{\beta ^2}4x^2SinF,
\eqno(2.6)
$$
where
$\beta =\frac{m_\pi }{eF_\pi }$,  $m_\pi $=140 MeV, e=4.84, $F_\pi $ =108 MeV.
	 
However, if one substitutes the kink-like profile function (2.5) in the left-hand side of (2.4) and compare with (2.6) (in Fig.1a and Fig.1b), it is seen that they are different.	  In order to make the equality to be satisfied approximately, we must modify the pion mass term  as following
$$
3.5 \times 10^{-7}\frac{\beta ^2}4x^2SinF.
\eqno(2.7)
$$
	
	After modification the pion mass term as (2.7), we have plotted it  in Fig.1c. We see that the approximate equality may be acceptable. So, one should be able to choice the symmetry-breaking term  as follow [6]
$$
{\cal L}_{SB}^{\left( mod.\right) }= \frac \varepsilon 8m_\pi ^2F_\pi ^2Tr\left( U+U^{+}-2\right) ,
\eqno(2.8)
$$
where $\varepsilon =3.5\times 10^{-7}$.

Now, we consider an ansazt has the form [13]
$$
N=\{Cosk\phi Sin\vartheta ,Sink\phi Sin\vartheta ,Cos\vartheta \},
\eqno(2.9)
$$
with this ansazt the soliton mass is

$$
M=M_2+M_4,
\eqno(2.10)
$$
$$
M_2=\frac \gamma 4\int_0^\infty dxx^2\int_0^\pi d\vartheta Sin\vartheta \{\left( F^{\prime }\right) ^2+[k^2+1]\frac{Sin^2F}{x^2}\},
\eqno(2.11)
$$
$$
M_4=\gamma \int_0^\infty dxx^2\int_0^\pi d\vartheta Sin\vartheta \{[\frac{Sin^2\vartheta }{Sin^2\vartheta }k^2+1]\left( F^{\prime }\right) ^2+\frac{Sin^2F}{x^2}k^2\}\frac{Sin^2F}{x^2},
\eqno(2.12)
$$
where
$
\gamma =\frac{\pi F_\pi }e, 
$
k is integer- k=1 corresponds to the case of the sherically symmetry hedgehod,  k$\geq 2$  to the case of twisty skyrmion configuration. The variation of  (2.10)  in F(x) give the following equation 
$$
[x^2+2a\sin ^2F]F^{^{\prime \prime }}+2xF^{^{\prime }}+[a(F^{^{\prime }})^2-\frac a4-2b\frac{\sin ^2F}{x^2}]\sin (2F)=0,
\eqno(2.13)
$$ 
where
$$
a=\int_0^\pi [k^2+1]\sin \vartheta d\vartheta ,
\eqno(2.14)
$$

$$
b=k^2\int_0^\pi \sin \vartheta d\vartheta, 
\eqno(2.15)
$$
which for $k=2$ we have $ a=10$ and $ b=8$.

From above-mentioned results and the formulas (2.11, 2.12), we consider the different cases of B=2 skyrmion configurations [6].

The first way of obtaining B=2 Skyrme field is to alter the boundary condition on the hedgehog profile function so that $F(0)=2\pi $  [14]. One can generate an approximation to this Skyrme field by using the kink-like profile function
$$
F\left( x\right) = 8arctan\left( e^{-x}\right) ,
\eqno(2.16)
$$
substitute it to (2.11) and (2.12) with k=1, we get
$$
M_a=M_2+M_4\simeq 46.7\gamma .
\eqno(2.17)
$$

	The second way of obtaining B=2 skyrme field is to leave the boundary condition on the profile unchanged but to have the skyrme field rotate twice as rapidly as the radial vector x under a change in the azimuthal angle around an axis [15]. From(2.1), (2.11) and (2.12) with k=2, we get
$$
M_b=M_2+M_4\simeq 38\gamma .
\eqno(2.18)
$$
	
	In the third case, based on the Lagrangian with ${\cal L}_{SB}$ (2.3) and $k=2$,  we obtained the equation of motion which is the equation (2.13) added by the term (2.6) in the right hand  side.
Solve this equation, we have nummerical dat file. Substituting it to (2.11) and (2.12) we get
$$
M_c=M_2+M_4\simeq 38.4\gamma .
\eqno(2.19)
$$

	Analogously, in the last case, based on the Lagrangian with 
${\cal L}_{SB}^{\left( mod.\right) }$ and $k=2$ we obtained the equation of motion which is the equation (2.13) added by the term (2.7) in the right hand  side.We find numerically F(x) (dat file) and substitute it to (2.11) and (2.12) we get
$$
M_d=M_2+M_4\simeq 27.5\gamma .
\eqno(2.20)
$$

\bigskip

\noindent
{\bf III. THE EQUATION FOR THE TIME -DEPENDENT HEDGEHOG

	AND THE TIME EVOLUTION OF THE SOLITON SOLUTION .}

\vspace{0.2cm}
We consider the Skyrme Lagrangian (2.1)  is added by a chiral symmetry-breaking term  (2.8).  It is convenient to parametrize the SU(2) matrix field U(x) by the pion field isovector  $\overrightarrow{\pi }\left(x\right) $ [16,17]

$$
U\left( x\right) =\frac{1+i\overrightarrow{\eta }\left( x\right) \overrightarrow{\tau }}{1-i\overrightarrow{\eta }\left( x\right) \overrightarrow{\tau }},
\eqno(3.1)
$$
where

$$\overrightarrow{\eta }\left( x\right) =\frac{\overrightarrow{\pi }\left( x\right) }{F_\pi },
\eqno(3.2)
$$
and $\vec \tau =(\tau _1,\tau _2,\tau _3)$ are the Pauli matrices.  In the parametrization (3.1) the expressions of the Cartan forms  $(\partial _\mu U)U^{+}$   for the SU(2) group have been calculated [16,17], and on the base of these expressions we get the expressions of the Lagrangian (2.1) anñ (2.8)  in the form

$${\cal L}=\frac{F_\pi ^2}2\frac{\left( \partial _\mu \overrightarrow{\eta }\partial ^\mu \overrightarrow{\eta }\right) }{\left( 1+\overrightarrow{\eta }^2\right) ^2}-\frac 4{e^2}\frac{\left[ \partial _\mu \overrightarrow{\eta }\times \partial ^\mu \overrightarrow{\eta }\right] ^2}{\left( 1+\overrightarrow{\eta }^2\right) ^4},
\eqno(3.3)
$$

$${\cal L}_{SB}^{\left( mod.\right) }=-\frac \varepsilon 2m_\pi ^2F_\pi ^2\frac{\overrightarrow{\eta }^2}{\left( 1+\overrightarrow{\eta }^2\right) }.
\eqno(3.4)
$$
The time-dependent hedgehog corresponds to the following ansatz

$$\overrightarrow{\eta }\left( r,t\right) =tan\left[ \frac{F\left( r,t\right) }2\right] \frac{\overrightarrow{r}}r,
\eqno(3.5)
$$
where F(r,t) is the profile function. Corresponding to this ansatz, ${\cal L}$ and ${\cal L}_{SB}^{\left( mod.\right) }$ are given by

$$
{\cal L}=\frac{F_\pi ^2}8\{(\dot F)^2-(F^{^{\prime }})^2-\frac{2\sin ^2F}{r^2}\}+\frac 1{2e^2}\frac{\sin ^2F}{r^2}\{2(\dot F)^2-2(F^{^{\prime }})^2-\frac{\sin ^2F}{r^2}\},
\eqno(3.6)
$$

$$
{\cal L}_{SB}^{\left( mod.\right) }=\frac{\varepsilon m_\pi ^2F_\pi ^2}4\left( \cos F-1\right). 
\eqno(3.7)
$$
The Hamiltonian is given by

$$
H=4\pi \int r^2dr\{\frac{F_\pi ^2}8\{(\dot F)^2+(F^{^{\prime }})^2+\frac{2\sin ^2F}{r^2}\}+\frac 1{2e^2}\frac{\sin ^2F}{r^2}\{2(\dot F)^2+2(F^{^{\prime }})^2+\frac{\sin ^2F}{r^2}\}+
$$
$$
\frac{\varepsilon m_\pi ^2F_\pi ^2}4\left( 1-\cos F\right) \}.
\eqno(3.8)
$$

The variational equation for the profile function is

$$
\frac{xF^{\prime }}2+\left[ \frac{x^2}4+2Sin^2F\right] (F^{^{\prime \prime }}-\ddot F)+Sin2F(F^{^{\prime }2}-\dot F^2)-\frac{Sin\left( 2F\right) }4-\frac{Sin^2FSin\left( 2F\right) }{x^2}
$$
$$
-\frac{\varepsilon \beta ^2}4x^2SinF=0,
\eqno(3.9)
$$
where 
$\beta =\frac{m_\pi }{eF_\pi }$,  $m_\pi $=140 MeV, e=4.84, $F_\pi $ =108 MeV [ 14 ] , and    $x=eF_\pi r$,  $\tau =eF_\pi t$  are the dimensionless distance and  time. The primes and the dots mean the derivatives with respect to x and $\tau$, respectively.  Hereafter  t is always understood as the dimensionless time.   

It is convenient to write F(x,t) in the form
$$
F\left( x,t\right) =F\left( x\right) +\delta F\left( x,t\right),
\eqno(3.10)
 $$
where F(x) is the profile function of the static hedgehog. To find $\delta F\left( x,t\right) $  we use the following boundary conditions 

$$\delta F\left( 0,t\right) =\delta F\left( L,t\right) =0.
\eqno(3.11)
$$
This condition is automatically satisfied by the harmonic expansion
$$\delta F\left( x,t\right) =\sum\limits_{j=1}^{N-1}A_j(t)\sin \left( \frac{j\pi x}L\right) ,
\eqno(3.12)
$$
where L is the size of  the spatial volume, N is the number of the points of the discretized spatial variable x, $A_j (t)$  are the amplitudes of jth fluctuation modes. These amplitudes are obtained by inverting the series given in (3.12)

$$
A_j(t)=\frac 2L\int_0^L\delta F(x,t)\sin (\frac{j\pi x}L)dx.
\eqno(3.13)
$$

 In our calculation we choose L=16, N=128 and the initial excitation mode is j =16. We denote $A\equiv A_{16}(0)$. We have studied the time evolution of the system for the pertubation parameter $A=0.1$.  The equation (3.9) should be reduced to a system coupled second order differential equations for the time variable t. We solve this system by using the Runge-Kutta procedure and obtain the solutions for $\delta F\left( x,t\right) $ and $A_j(t)$.  We have plotted the fluctuation  $\delta F\left( x,t\right) $ at t=0, 100, 200, 300 and 500  in Fig.2a to Fig.2e, respectively. We see that apart from large fluctuations near $x=0$, $\left| \delta F(x,t)\right| \sim 0.1$  . This is understandable as the Skyrmion dynamics is dominated by the small x-region and the deviation from the Skyrmion is small. The dynamical behavior of the system is understood better by observing the time evolution of the amplitudes of various harmonic modes.  In Fig.3a to Fig.3c we have plotted $A_j(t)$ for  j=8, 16, 32 , respectively.  It is seen that  the amplitude of the initial mode decreases gradually and it has some kind of periodicity in the variation,  while amplitudes of other modes increase on the average. The time evolution of the mode of j=8 and j=16 has a periodic behavior  when it was consider in a small interval of the time.  It is to be expected that after a longer interval of time, all the modes would be of the same magnitude leading to "thermalization" and "spatio-temporal chaos". 

\bigskip

\noindent
{\bf   IV.  DYNAMICAL BEHAVIOR  OF SKYRMION  IN  A GENERALIZED  SKYRME  MODEL .}


We consider  the  Lagrangian 

$$
{\cal L} = {\cal L}^{(2)} + {\cal L}^{(4)} + {\cal L}^{(6)} +  {\cal L}_{SB}^{\left( mod.\right) },
\eqno(4.1)
$$
where  ${\cal L}^{(2)} +  {\cal L}^{(4)} $  is the Lagrangian of the Skyrme model (2.1),  ${\cal L}^{(6)}$ is  given by  (1.2 ),  ${\cal L}_{SB}^{\left( mod.\right) }$   is  given by (2.8).   In the parametrization (3.1), the expressions of the Cartan forms  $(\partial _\mu U)U^{+}$   for the SU(2) group have been calculated [16,17].  We get  the Lagrangian (1.2)  in the form

$$
{\cal L}^{(6)}=\frac{4\varepsilon _6^2}{3\pi ^4}\frac 1{\left( 1+\overrightarrow{\eta }^2\right) ^6}\{(\partial \overrightarrow{\eta })^6-3(\partial \overrightarrow{\eta })^2(\partial _\alpha \overrightarrow{\eta }\partial ^\beta \overrightarrow{\eta })(\partial _\beta \overrightarrow{\eta }\partial ^\alpha \overrightarrow{\eta })+
$$

$$
2(\partial _\nu \overrightarrow{\eta }\partial ^\beta \overrightarrow{\eta })(\partial _\beta \overrightarrow{\eta }\partial ^\alpha \overrightarrow{\eta })(\partial _\alpha \overrightarrow{\eta }\partial ^\nu \overrightarrow{\eta })\}.
\eqno(4.2)
$$
By the same  caculation in Sec. III, we have obtained the  variational equation for the profile function
$$
(\frac{x^2}4+2Sin^2F+\frac \gamma 4\frac{Sin^4F}{x^2})(\ddot F-F^{"})+(Sin\left( 2F\right) +\frac \gamma 4\frac{Sin^2FSin\left( 2F\right) }{x^2})(\dot F^2-F^{\prime 2})-
$$
$$
(\frac x2-\frac \gamma 4\frac{Sin^4F}{x^3})F^{\prime }+\frac{Sin\left( 2F\right) }4+\frac{Sin^2FSin\left( 2F\right) }{x^2}+-\frac{\varepsilon \beta ^2}4x^2SinF = 0,
\eqno(4.3)
$$
where  $x=eF_\pi r$,  $\tau =eF_\pi t$  are the dimensionless distance and  time. The primes and the dots mean the derivatives with respect to x and $\tau$, respectively,  $\gamma =\frac{F_\pi ^2\varepsilon _6^2e^4}{\pi ^4}$,  $\beta =\frac{m_\pi }{eF_\pi }$. We choose   $\varepsilon _6^2=5fm^2$,  and  t is  understood as the dimensionless time. 

Based on the expressions  (3.10) to (3.13)  and the choice of L=16, N=128, $A\equiv A_{16}(0)=0.1$, we consider the dynamical behavior of the system with Lagragian ( 4.1) by observing the time evolution of the amplitudes of various harmonic modes.  We have plotted the  profile function of the static hedgehog  and the  fluctuation  $\delta F\left( x,t\right) $ at  t = 0, 100, 200, 300 in Fig. 4a to Fig. 4d.   It is a clear indication that   $\delta F\left( x,t\right) $  has violent  fluctuations  for substantially longer intervalls in x. 

The plots of  $A_j(t)$ for the cases of  j= 8, 16, 64, 127  ( Fig. 5a to Fig. 5d )  indicate that the fluctuation amplitudes in the generalized Skyrme model which Lagrangian added by the sixth-order term   develop  much faster than in the Skyrme model.  For this model, the amplitudes of   "spontaneous" fluctuations appear after  $t\approx 150$.
One can say that a self-exitation of soliton takes place after  $t\approx 150$.

\bigskip


{{\bf V. CONCLUSION}}


\vspace{0.2cm}
  
   We have calculated the energy of four B=2 skyrmion configurations. The comparison of the values of the energy in four cases shows that in the case of modified Lagrangian one could obtain B=2 configuration with lowest energy. That is, this configuration is more close to the real energy minimum of B=2 skyrmion.

The B=1 hedgehog kink-like profile function (2.5) is a convenient analytic approximation to the numerical solution. But how can modify the Lagrangian to obtain the profile function (2.5), the modification made in this paper is one of the answers to the question. 

By integrating the equation (3.1) for the time-dependent hedgehog we have obtained the information about the dynamical behavior of the sotiton in the model (3.3, 3.4). We have plotted $\delta F\left( x,t\right) $ at various moments and the development of amplitudes of fluctuation modes $ A_j(t)$. The plots of  $A_j(t)$  for the case A=0.1 indicate that the process of thermalization takes place sooner, and the fluctuation amplitudes in the considered model  develop much faster  than in the Skyrme model (see [7]). Besides, in [7], for A=0.1 the amplitude of the initial (32nd) mode has extremely regular periodic behavior while in our initial (16nd) mode the amplitudes decreases gradually on the average.

When higher order terms are included, the theory becomes much more nonlinear, so it is clearly  that the phase space around the soliton solution  in the generalized Skyrme model was shown to be more stochastic than in the original Skyrme model.

We are grateful to To Ba Ha for valuable help.
This  work was supported in part by the National Basic Research Program in Natural Sciences under the grant number KT-04-1.2/99.
\bigskip

\noindent
{{\bf REFERENCES    }}

\vspace{0.2cm} 

\noindent
1.  Skyrme T.H.R. , {\it Proc.Roy.Soc.}, {\bf A260} (1961) 127; Nucl.Phys. {\bf 31} (1962) 556. 

 \noindent
2.  Adkins G. , Nappi C. , Witten E. , {\it Nucl.Phys.}, {\bf B228} ( 1983) 552.

\noindent
3.  Nikolaev B.A. , {\it /Fiz.Elem.Chatstists.At.Yadra.}, {\bf 20} (1989)  420 (in Russian).

\noindent
4.  Atiyal M.I. , Manton N.S. , {\it Phys.Lett.}, {\bf B222} (1989) 438.

\noindent
5.  Sutcliffe P.M. , {\it Phys.Lett.}, {\bf B292} (1992) 104.

\noindent
6. Nguyen Vien Tho, Le Trong Tuong, Phu Chi Hoa,  {\it Comm. in Phys.}, {\bf Vol.9} (1999) 73-77.

\noindent
7.   Segar J. ,  Siram V. , {\it Phys. Rev.}, {\bf  D53} (1996) 3876.

\noindent
8.   Jackson A. ,  Jackson A.  D. , Goldhaber A. S. , Brown G. E . ,{\it  Phys.  Lett.}, {\bf  B154} (1985) 101.

\noindent
9.   Wirzba A. ,  Weise W. ,{\it  Phys.  Lett.}, {\bf  B188} (1987) 6.

\noindent
10.  Dube S. , Marleau L. ,{\it  Phys.  Rev.}, {\bf  D41} (1990) 1606.

\noindent
11.    Marleau L. , {\it  ibid.}, {\bf  D43} (1991) 885.

\noindent
12.    Jackson A.  D. , Weiss C. , Wirzba A. ,{\it  Nucl.  Phys. }, {\bf  A529} (1991) 741.

\noindent
13.   Nikolaiev B. A.  ,  Tkachev O. F. , {\it Phys. Elem. Chastic}, {\bf Tom21} (1990). 

\noindent
14.   Jackson A. D.  , {\it  Pro. Phys. Rev. Lett.}, {\bf 51} (1983) 751.

\noindent
15.  Weigel  H.  , Schwesinger  B. and  Holzwarth G. , {\it Phys. Lett.}, {\bf B168} (1986) 321.

\noindent
16.  Kushinov V. I. , Nguyen Vien Tho, {\it Fiz. Elem. Chatstists. At. Yadra.}, {\bf Vol 25} (1994) 603.  

\noindent
17.  Nguyen Vien Tho, {\it Comm. in Phys.}, {\bf Vol.7} (1997) 1-9.

\end{document}